\documentclass[]{interact}

\usepackage{epstopdf}
\usepackage{subfigure}
\usepackage{hyperref,bm,amsmath,booktabs,multirow,color}

\usepackage{natbib}
\bibpunct[, ]{(}{)}{;}{a}{}{,}
\renewcommand\bibfont{\fontsize{10}{12}\selectfont}

\theoremstyle{plain}

\theoremstyle{definition}

\theoremstyle{remark}

\begin{document}

\articletype{Original Paper}

\title{Variable fusion for Bayesian linear regression via spike-and-slab priors}

\author{
\name{Shengyi Wu\textsuperscript{a}, Kaito Shimamura\textsuperscript{a}, Kohei Yoshikawa\textsuperscript{a}, Kazuaki Murayama\textsuperscript{a} and Shuichi Kawano\textsuperscript{a}\thanks{CONTACT Shuichi Kawano. Email: skawano@ai.lab.uec.ac.jp}}
\affil{\textsuperscript{a}Graduate School of Informatics and Engineering, The University of Electro-Communications, 1-5-1 Chofugaoka, Chofu-shi, Tokyo 182-8585, Japan.}
}

\maketitle

\begin{abstract}
In linear regression models, fusion of coefficients is used to identify predictors having similar relationships with a response. 
This is called variable fusion. 
This paper presents a novel variable fusion method in terms of Bayesian linear regression models. 
We focus on hierarchical Bayesian models based on a spike-and-slab prior approach. 
A spike-and-slab prior is tailored to perform variable fusion. 
To obtain estimates of the parameters, we develop a Gibbs sampler for the parameters. 
Simulation studies and a real data analysis show that our proposed method achieves better performance than previous methods.
\end{abstract}


\section{Introduction}
\label{sec:Introduction}
In recent years, because of the rapid development of computer hardware and systems, a wide variety of data are being observed and recorded in genomics, medical science, finance, and many other fields of science. 
Linear regression is a fundamental statistical method for extracting useful information from such datasets. 
In linear regression models, fusion of coefficients is used to identify predictors having similar relationships with a response. 
This is called variable fusion \citep{land1996variable}. 
Many kinds of research on variable fusion have been conducted to date. 
For example, we refer the reader to \cite{tibshirani2005sparsity,tibshirani2007spatial,bondell2008simultaneous,kim2009ell_1,tibshirani2011solution,ma2017concave}. 
In addition, there has been much research on variable fusion in areas besides linear regression \citep{HockingVBclusterpathJ,danaher2014joint,molstad2019penalized,price2019automatic,dondelinger2020joint}. 
Variable fusion is essentially achieved by modifying regularization terms that perform variable selection (e.g., the lasso \citep{tibshirani1996regression}): the fused lasso \citep{tibshirani2005sparsity}, the OSCAR \citep{bondell2008simultaneous}, the clustered lasso \citep{she2010sparse}, and so on. 
It can be regarded as a frequentist approach.


On the other hand, only a few methods for variable fusion have been reported in terms of a Bayesian approach. 
The Bayesian approach is based on priors that induce variable selection: the Laplace prior \citep{williams1995bayesian,park2008bayesian}, the normal-exponential-gamma (NEG) prior \citep{griffin2005alternative}, the horseshoe prior \citep{carvalho2010horseshoe}, the Dirichlet-Laplace prior \citep{bhattacharya2015dirichlet}, and the spike-and-slab prior \citep{george1993variable,ishwaran2005spike,fruhwirth2010bayesian}. 
For example, the Bayesian fused lasso by \cite{kyung2010penalized} is based on the Laplace prior, and the Bayesian fusion method by \cite{shimamura2019bayesian} is based on the NEG prior. 
The spike-and-slab prior is often used in the context of Bayesian variable selection. 
However, there are few research studies about spike-and-slab priors that include variable fusion. 

In this paper, we propose a variable fusion method in the framework of Bayesian linear regression with a spike-and-slab prior. 
The spike-and-slab prior is based on the Dirac spike prior \citep{fruhwirth2010bayesian} and the $g$-slab prior \citep{zellner1986bayesian}. 
We tailor the Dirac spike prior and the $g$-slab prior to perform variable fusion by assuming priors on the difference between adjacent parameters. 
We adopt hierarchical Bayesian models based on the spike-and-slab prior with variable fusion. 
To obtain estimates of the parameters, we develop a Gibbs sampler.

 
The remainder of the paper is organized as follows. 
Section \ref{sec:BayesianVRviaSSP} describes Bayesian linear regression and Bayesian variable selection with spike-and-slab priors. 
In Section \ref{sec:ProposedMethod}, we introduce a spike-and-slab prior that performs variable fusion and then build a Bayesian linear regression model based on the prior. 
In addition, we develop a Gibbs sampling method to obtain estimates of the parameters. 
Section \ref{sec:NumericalStudies} reports Monte Carlo simulations and real data analysis conducted to examine the performance of our proposed method and to compare it with previous methods. 
Conclusions are given in Section \ref{sec:Conclusions}.

\section{Bayesian variable selection via spike-and-slab priors}
\label{sec:BayesianVRviaSSP}
\subsection{Bayesian linear regression}
\label{sec:BayesianLR}
We consider the following linear regression model:
\begin{equation}
\bm{y}=\bm{X}\bm{\beta} +\bm{\varepsilon},
\label{eq:LinearRegression}
\end{equation}
where $\bm{y}={({y}_{1},\ldots,{y}_{n})}^{T}$ is a vector of the response variable, $\bm{X}=({\bm{x}}_{(1)},\ldots,{\bm{x}}_{(p)})$ is a design matrix of the predictors,   $\bm{\beta}={({\beta}_{1},\ldots,{\beta}_{p})}^{T}$ comprises the coefficients to be estimated, and $\bm{\varepsilon}={({\varepsilon}_{1},\ldots,{\varepsilon}_{n})}^{T}$ is a vector of errors with zero means and variance-covariance matrix $\sigma^2 {\bm I}_n$. 
Here, ${\bm{x}}_{(j)}={({x}_{1j},\ldots ,{x}_{nj})}^{T}$, $\sigma^2 \ (\sigma>0)$ is the variance of an error, and $\bm{{I}}_{n}$ is the $n \times n$ identity matrix. 
In addition, the response is centered and the predictors are standardized as follows:
\begin{equation*}
\sum_{i=1}^{n}{y}_{i}=0,\quad \sum_{i=1}^{n}{x}_{ij}=0, \quad \sum_{i=1}^{n}{x}^{2}_{ij}=n,\quad(j=1,\ldots,p).
\end{equation*}
The centering and standardization allow us to omit the intercept from Equation \eqref{eq:LinearRegression}.

We assume that the error vector $\bm{\varepsilon}$ is distributed as ${{\rm N}}_{n}({\bm{0}}_{n},{\sigma}^{2}\bm{{I}}_{n})$, where ${\bm 0}_n$ is the $n$-dimensional zero vector. 
Then, the likelihood is given by
\begin{equation}
	p(\bm{y}|\bm{X};\bm{\beta},{\sigma}^{2})=\prod_{i=1}^{n}p({y}_{i}|{\bm{x}}_{i};\bm{\beta},{\sigma}^{2}),
	\label{eq:likelihood}
\end{equation}
where ${\bm{x}}_{i}=({x}_{i1},\ldots,{x}_{ip})$ and
\begin{equation*}
	p({y}_{i}|{\bm{x}}_{i};\bm{\beta},{\sigma}^{2})=\frac{1}{\sqrt{2\pi{\sigma}^{2}}}\exp \left\{-\frac{{({y}_{i}-{\bm{x}}^{T}_{i}\bm{\beta})}^{2}}{2{\sigma}^{2}} \right\}.
\end{equation*}
Hereinafter, the likelihood $p(\bm{y}|\bm{X};\bm{\beta},{\sigma}^{2})$ will be denoted as $p(\bm{y}|\bm{\beta},{\sigma}^{2})$ for simplicity.

A Bayesian linear regression model is formed by the likelihood and priors of parameters $\bm{\beta}$ and ${\sigma}^{2}$. 
As a prior of the parameter $\bm{\beta}$, a conjugated prior
\begin{equation}
	\bm{\beta}\sim {\rm N}_{p}({\bm{b}}_{0},{\bm{B}}_{0})
	\label{eq:conjugateprior}
\end{equation}
is often assumed. 
Here, ${\bm{b}}_{0}$ is a $p$-dimensional vector and ${\bm{B}}_{0}$ is a $p \times p$ symmetric matrix. 
Both are hyper-parameters. 
Also, we often assume an inverse Gamma distribution or a uniform distribution for the prior of the parameter $\sigma^2$. 
Although it is difficult to obtain posteriors of the parameters $\bm{\beta}$ and ${\sigma}^{2}$ from the likelihood \eqref{eq:likelihood} and the prior \eqref{eq:conjugateprior}, we can obtain the full conditional posteriors of the parameters. 
Using a Gibbs sampler from the full conditional posteriors, we can infer the parameters. 
We refer the reader to \cite{gelman2013bayesian} for details of the full conditional posteriors and the Gibbs sampler.

\subsection{Spike-and-slab priors}
\label{sec:SSP}
The prior \eqref{eq:conjugateprior} does not induce variable selection. 
This means that none of the components of the coefficients are estimated as being exactly zero. 
Many researchers have studied priors to perform variable selection, as mentioned in the Introduction. 
We focus on the spike-and-slab prior, given by
\begin{equation}
	p(\bm{\beta}|\bm{\xi})={p}_{\rm slab}({\bm{\beta}}_{\bm{\xi}}|\bm{\xi})\prod^{}_{j:{\xi}_{j}=0}{p}_{\rm spike}({\beta}_{j}|{\xi}_{j}),
	\label{eq:sps}
\end{equation}
where $p_{\rm slab} (\cdot)$ is a slab prior,  $p_{\rm spike} (\cdot)$ is a spike prior, ${\bm \xi}=(\xi_1,\ldots,\xi_p)$ is a vector of latent indicator variables that take value zero or one, and ${\bm \beta}_{\bm \xi}$ the vector comprising the elements of $\bm \beta$ for which $\xi_j=1$. 
The $j$-th parameter $\beta_j$ belongs to the slab prior or spike prior when $\xi_j=1$ or $\xi_j=0$, respectively. 
The slab prior has its mass spread over the wide range of possible values for the coefficients. 
On the other hand, the mass of the spike prior is concentrated around zero, encouraging variable selection. 
Having the mass concentrated around zero means that the value of the corresponding coefficient will be zero.

As the slab prior, we often assume a multivariate normal distribution with mean vector ${\bm b}_{0, \bm \xi}$ and variance-covariance matrix $\sigma^2 {\bm B}_{0, \bm \xi}$. 
Specifically, the following three types of slab priors are widely used:
%
\begin{itemize} 
\item The independence slab (i-slab): ${\bm{b}}_{0,\bm{\xi}}=\bm{0}_{p_0}$ and ${\bm{B}}_{0,\bm{\xi}}=c\bm{I}_{p_0}$, 
\item The g-slab: ${\bm{b}}_{0,\bm{\xi}}=\bm{0}_{p_0}$ and ${\bm{B}}_{0,\bm{\xi}}=g{({\bm{X}}^{T}_{\bm{\xi}}{\bm{X}}_{\bm{\xi}})}^{-1}$, 
\item The fractional slab (f-slab): ${\bm{b}}_{0,\bm{\xi}}={({\bm{X}}^{T}_{\bm{\xi}}{\bm{X}}_{\bm{\xi}})}^{-1}{\bm{X}}^{T}_{\bm{\xi}}\bm{y}$ and ${\bm{B}}_{0,\bm{\xi}}=\frac{1}{b}{({\bm{X}}^{T}_{\bm{\xi}}{\bm{X}}_{\bm{\xi}})}^{-1}$, 
\end{itemize}
where $c$, $g$, and $b$ are positive hyper-parameters, ${p}_{0}=\sum_{j=1}^p{\xi}_{j}$, ${\bm{X}}_{\bm{\xi}}$ is the design matrix consisting of columns of $\bm{X}$ corresponding to the coefficients allocated to the slab prior.
Note that the g-slab is the Zellner's $g$-prior introduced by \cite{zellner1986bayesian} and the f-slab is the fractional prior introduced by \cite{o1995fractional}.

There are two types of spike priors: a Dirac spike and an absolutely continuous spike. 
The Dirac spike was proposed by \cite{fruhwirth2010bayesian}. 
It can be specified as ${p}_{\rm spike}({\beta}_{j}|{\xi}_{j})={\Delta}_{0}({\beta}_{j})$, where ${\Delta}_{0}(\cdot)$ is the Dirac measure defined by
\begin{equation*}
{\Delta}_{x_0}(x) = \begin{cases}
    0, \ (x \neq x_0), \\
    1, \ (x = x_0).
  \end{cases}
\end{equation*}
The Dirac spike can set the corresponding coefficients to be exactly zero. 
For the absolutely continuous spike, we refer the reader to \cite{george1993variable,ishwaran2003detecting,ishwaran2005spike}.

To obtain the Gibbs sampler from the likelihood \eqref{eq:likelihood} and the spike-and-slab prior \eqref{eq:sps}, we impose a prior on the latent indicator variables $\xi_j$ hierarchically in the form
\begin{equation*}
	p({\xi}_{j}|\omega)={\omega}^{{\xi}_{j}}{(1-\omega)}^{1-{\xi}_{j}}, \quad
	\omega \sim  {\rm Beta}{({a}_{\omega},{b}_{\omega})},
\end{equation*}
where $\omega$ is the probability of ${\xi}_{j}=1$, $\rm{Beta}(\cdot,\cdot)$ is the beta distribution, and ${a}_{\omega}$ and ${b}_{\omega}$ are positive hyper-parameters. 
If we adopt a Dirac spike as the spike prior and impose an uninformative prior $p({\sigma}^{2})=1/{\sigma}^{2}$ on ${\sigma}^{2}$, we can derive the full conditional posteriors of the parameters as follows. 
The full conditional posterior of ${\xi}_{j}$ is given by
\begin{align*}
	{\xi}_{j}|\bm{y},\omega,{\bm{\xi}}_{-j} &\sim {\left( \frac{\omega p(\bm{y}|{\xi}_{j}=1,{\bm{\xi}}_{-j})}{\omega p(\bm{y}|{\xi}_{j}=1,{\bm{\xi}}_{-j})+(1-\omega) p(\bm{y}|{\xi}_{j}=0,{\bm{\xi}}_{-j})} \right)}^{{\xi}_{j}}\\
	&\quad \times{\left(\frac{\omega p(\bm{y}|{\xi}_{j}=0,{\bm{\xi}}_{-j})}{\omega p(\bm{y}|{\xi}_{j}=1,{\bm{\xi}}_{-j})+(1-\omega) p(\bm{y}|{\xi}_{j}=0,{\bm{\xi}}_{-j})}\right)}^{1-{\xi}_{j}},
\end{align*}
where ${\bm{\xi}}_{-j}$ denotes the vector consisting of all the elements of $\bm{\xi}$ except ${\xi}_{j}$, and $p(\bm{y}|\bm{\xi})$ is given by
\begin{equation*}
	p(\bm{y}|\bm{\xi})= \frac { 1 }{ { (2\pi ) }^{ n/2 } } \frac { { |{ \bm{B} }_{  \bm{\xi}   }| }^{ 1/2 } }{ { |{ \bm{B} }_{ 0,\bm{\xi} }| }^{ 1/2 } }  \frac { \Gamma (n/2) }{ { {s}_{n} }^{ n/2 } }.
	\end{equation*}	
Here $\bm{B}_{\bm{\xi}}={({\bm{X}}^{T}_{\bm{\xi}}{\bm{X}}_{\bm{\xi}}+\frac{1}{c}\bm{I}_{p_0})}^{-1}$ if we use the i-slab, $\bm{B}_{\bm{\xi}}=\frac{g}{g+1}{({\bm{X}}^{T}_{\bm{\xi}}{\bm{X}}_{\bm{\xi}})}^{-1}$ if we use the g-slab, $\bm{B}_{\bm{\xi}}={({\bm{X}}^{T}_{\bm{\xi}}{\bm{X}}_{\bm{\xi}})}^{-1}$ if we use the f-slab, $\Gamma(\cdot)$ is the Gamma function, and ${s}_{n}=({\bm{y}}^{T}\bm{y}-{\bm{b}}^{T}_{\bm{\xi}}{\bm{B}}^{-1}_{\bm{\xi}}{\bm{b}}_{\bm{\xi}})/2$.  
The full conditional posteriors of the other parameters are then given as follows:
\begin{align*}
    \omega|\bm{\xi} &\sim {\rm Beta}({a}_{\omega}+{p}_{0},{b}_{\omega}+p-{p}_{0}),\\
	{\sigma}^{2}|\bm{y},\bm{\beta},\bm{\xi}&\sim {\rm IG}\left(\frac{n}{2},\frac{{\bm{y}}^{T}\bm{y}-{\bm{b}}^{T}_{\bm{\xi}}{\bm{B}}^{-1}_{\bm{\xi}}{\bm{b}}_{\bm{\xi}}}{2}\right),\\
	{\beta}_{j}|{\xi}_{j}&=0\quad(j:{\xi}_{j}=0),\\
	{\bm{\beta}}_{\bm{\xi}}|\bm{y},\bm{\xi},{\sigma}^{2}&\sim {\rm N}_{p_0}({\bm{b}}_{\bm{\xi}},{\sigma}^{2}{\bm{B}}_{\bm{\xi}})\quad(j:{\xi}_{j}=1),
\end{align*}
where ${\rm IG} (\cdot, \cdot)$ is an inverse Gamma distribution and ${\bm{b}}_{\bm{\xi}}={\bm{B}}_{\bm{\xi}}{\bm{X}}^{T}_{\bm{\xi}}\bm{y}$.

\section{Proposed method}
\label{sec:ProposedMethod}
In this section, we propose a Bayesian linear regression with variable fusion. 
Our proposed method is based on spike-and-slab priors that fuse adjacent coefficients in linear regression models. 
First, we introduce spike-and-slab priors that perform variable fusion, and then we derive a Bayesian model based on the priors. 
A Gibbs sampler is also provided for the estimates of the parameters.

\subsection{Fusing adjacent coefficients by spike-and-slab priors}
\label{sec:Fusing}

We consider the likelihood \eqref{eq:likelihood}. 
Let a vector $\bm{\gamma}={({\gamma}_{1},\ldots,{\gamma}_{p-1})}^{T}$ be the differences between adjacent elements of the coefficients $\bm{\beta}$; that is, ${\gamma}_{j}={\beta}_{j+1}-{\beta}_{j} \ (j=1,\ldots,p-1)$. 
We assume that $\bm{\gamma}$ follows the spike-and-slab prior
\begin{equation}
	p(\bm{\gamma}|\bm{\delta})={p}_{\rm slab}({\bm{\gamma}}_{\bm{\delta}}|\bm{\delta})\prod^{}_{j:{\delta}_{j}=0}{p}_{\rm spike}({\gamma}_{j}|{\delta}_{j}),
	\label{eq:sps_fusion}
\end{equation}
where $\bm{\delta}=({\delta}_{1},\ldots,{\delta}_{p-1})$ comprises latent indicator variables having value zero or one, and ${\bm{\gamma}}_{\bm{\delta}}$ comprises the elements of $\bm{\gamma}$ corresponding to ${\delta}_{j}=1$. 
The prior \eqref{eq:sps_fusion} indicates that when ${\delta}_{j}=0$, the corresponding ${\gamma}_{j}$ is allocated to the spike component. 
The other elements of $\bm{\gamma}$ (i.e., those with ${\delta}_{j}=1$) are allocated to the slab component.

We adopt the Dirac spike and the g-slab prior. 
The reason for this choice is that \cite{malsiner2011comparing} showed through various simulations that the Dirac spike combined with g-slab has the best performance among five types of spike-and-slab priors with respect to encouraging sparsity. 
Therefore, we set ${p}_{\rm spike}({\gamma}_{j}|{\delta}_{j})={\Delta}_{0}({\gamma}_{j})$. 
Thus, when ${\delta}_{j}=0$, the corresponding ${\gamma}_{j}$ is set to zero. 
This implies ${\beta}_{j+1}={\beta}_{j}$. 
Also, we set ${p}_{\rm slab}({\bm{\gamma}}_{\bm{\delta}}|\bm{\delta})={\rm N}_{p_1}({\bm 0}_{p_1},{\sigma}^{2}{\bm{H}}_{0,\bm{\delta}})$, where ${p}_{1}=\sum_{j=1}^{p-1} {\delta}_{j}$.

Before specifying the variance-covariance matrix ${\bm{H}}_{0,\bm{\delta}}$ in ${p}_{\rm slab}({\bm{\gamma}}_{\bm{\delta}}|\bm{\delta})$, we need to define ${\bm \beta}_{\bm \delta}$ and specify its prior. 
We define an $n \times (p_1+1)$ matrix ${\bm{X}}_{\bm{\delta}}$ whose columns are composed by ${\bm{x}}_{(j)}+{\bm{x}}_{(j+1)}$ for the set $\{ j | {\delta}_{j}=0 \}$ and ${\bm{x}}_{(j)}$ for the set $\{ j | {\delta}_{j}=1 \}$. 
Let ${\bm{\beta}}_{\bm{\delta}}$ be a vector of the elements of $\bm{\beta}$ corresponding to ${\bm{X}}_{\bm{\delta}}$. 
Furthermore, we assume the $g$-prior for ${\bm{\beta}}_{\bm{\delta}}$ in the form ${\rm N}_{p_1+1}({\bm 0}_{p_1+1},{\sigma}^{2}{\bm{B}}_{0,\bm{\delta}})$, where ${\bm{B}}_{0,\bm{\delta}}=g{({\bm{X}}^{T}_{\bm{\delta}}{\bm{X}}_{\bm{\delta}})}^{-1}$ and $g$ is a positive hyper-parameter. 

%

Using the variance-covariance matrix ${\bm B}_{0, \bm \delta}$, we can obtain the detailed formulation of the variance-covariance matrix ${\bm H}_{0, \bm \delta}$. 
We set $z_{ij} \ (i,j=1,\ldots,p_1+1)$ equal to the $(i,j)$-th element of ${\bm B}_{0, \bm \delta}$ and $\alpha_{ij} \ (i,j=1,\ldots,p_1)$ equal to the $(i,j)$-th element of  ${\bm H}_{0, \bm \delta}$.
Then $\alpha_{ij}$ is can be written as
%
\begin{equation*}
	{\alpha}_{ij}={z}_{(i+1)(j+1)}-{z}_{(i+1)j}-{z}_{i(j+1)}+{z}_{ij}\quad(i,j=1,\ldots,p_1).
\end{equation*}
According to the above procedure, we can specify the slab prior ${p}_{\rm slab}({\bm{\gamma}}_{\bm{\delta}}|\bm{\delta})={\rm N}_{p_1}({\bm 0}_{p_1},{\sigma}^{2}{\bm{H}}_{0,\bm{\delta}})$. 

By specifying the priors for $\delta_j$ similar as in Section \ref{sec:SSP} and an uninformative prior $p({\sigma}^{2})=1/{\sigma}^{2}$ on ${\sigma}^{2}$, the proposed Bayesian hierarchical model is given by
\begin{align*}
	\bm{y}|\bm{\beta},{\sigma}^{2} &\sim  {{\rm N}}_{n}(\bm{X}\bm{\beta},{\sigma}^{2}{\bm{I}}_{n}),\\
	{\gamma}_{j} \ (={\beta}_{j+1}-{\beta}_{j})|{\delta}_{j}&=0\quad(j:{\delta}_{j}=0),\\
	{\bm{\gamma}}_{\bm{\delta}} \ (={\bm{D}}_{\bm{\beta}}{\bm{\beta}}_{\bm{\delta}})|\bm{\delta},{\sigma}^{2} &\sim  {\rm N}_{p_1}({\bm{h}}_{0,\bm{\delta}},{\sigma}^{2}{\bm{H}}_{0,\bm{\delta}})\quad(j:{\delta}_{j}=1),\\
	{\delta}_{j}|\omega &\sim {\omega}^{{\delta}_{j}}{(1-\omega)}^{1-{\delta}_{j}},\\
	\omega &\sim  {\rm Beta}({a}_{\omega},{b}_{\omega}),\\
	{\sigma}^{2} &\sim  \frac{1}{{\sigma}^{2}},
\end{align*}
where ${\bm{D}}_{\bm{\beta}}$ is the ${p}_{1} \times ({p}_{1}+1)$ matrix defined by
\begin{eqnarray*}
	{\bm{D}}_{\bm{\beta}}=\begin{pmatrix} -1 & 1 & 0 & 0 & \cdots  & 0 & 0 \\ 0 & -1 & 1 & 0 & \cdots  & 0 & 0 \\ \vdots  & \vdots  & \vdots  & \vdots  & \ddots  & \vdots  & \vdots  \\ 0 & 0 & 0 & 0 & \cdots  & -1 & 1 \end{pmatrix}.
\end{eqnarray*}


\subsection{Full conditional posteriors}
\label{sec:FullConditional}
It is feasible to infer the posterior by using the Markov chain Monte Carlo (MCMC) method: the model parameters $(\bm{\delta},\omega,{\sigma}^{2},\bm{\beta})$ are sampled from their full conditional posteriors.

In building the Gibbs sampler of the parameters, it is essential to draw $\bm{\delta}$ from the marginal posterior
\begin{equation*}
	p(\bm{\delta}|\bm{y})\propto p(\bm{y}|\bm{\delta})p(\bm{\delta}),
\end{equation*}
where $p(\bm{y}|\bm{\delta})$ is the marginal likelihood of the linear regression model with ${\bm{X}}_{\bm{\delta}}$. 
This marginal likelihood can be derived analytically as
\begin{equation}
	p(\bm{y}|\bm{\delta})= \frac { 1 }{ { (2\pi ) }^{ (n-1)/2 } } \frac { { |{ \bm{H} }_{ { \bm{\delta}  } }| }^{ 1/2 } }{ { |{ \bm{H} }_{ 0,\bm{\delta} }| }^{ 1/2 } }  \frac { \Gamma (n/2) }{ { {s}_{c} }^{ n/2 } }.
	\label{eq:ML}
	\end{equation}
Here, ${\bm{H}}_{\bm{\delta}}$ and ${s}_{c}$ are respectively given by
\begin{align*}
	{\bm{H}}_{\bm{\delta}}&={ ({ \bm{X} }^{ T }_{\bm{\delta}}{\bm{X}}_{\bm{\delta}}+{ {\bm{D}}_{\bm{\delta}} }^{ T }{ \bm{H} }_{ 0 ,\bm{\delta}}^{ -1 }{\bm{D}}_{\bm{\delta}}) }^{ -1 },\\
	{s}_{c}&=\frac { 1 }{ 2 } ({ \bm{y} }^{ T }\bm{y}+{ \bm{h} }_{ 0 ,\bm{\delta}}^{ T }{ \bm{H} }_{ 0 ,\bm{\delta}}^{ -1 }{ \bm{h} }_{ 0 ,\bm{\delta}}-{ \bm{h} }_{ \bm{\delta}  }^{ T }{ \bm{H} }_{ \bm{\delta}  }^{ -1 }{ \bm{h} }_{ \bm{\delta}  }),
\end{align*}
where ${\bm{h}}_{\bm{\delta}} = { \bm{H}}_{ \bm{\delta}  }({ \bm{X} }^{ T }_{\bm{\delta}}\bm{y}+{ \bm{D} }^{ T }_{\bm{\delta}}{ \bm{H} }_{ 0 ,\bm{\delta}}^{ -1 }{ \bm{h} }_{ 0 ,\bm{\delta}})$.
The derivation of the marginal likelihood is given in Appendix A. 

Thus the full conditional posterior of ${\delta}_{j}$ is given by
\begin{align*}
	{\delta}_{j}|\bm{y},\omega,{\bm{\delta}}_{-j} &\sim {\Big(\frac{\omega p(\bm{y}|{\delta}_{j}=1,{\bm{\delta}}_{-j})}{\omega p(\bm{y}|{\delta}_{j}=1,{\bm{\delta}}_{-j})+(1-\omega) p(\bm{y}|{\delta}_{j}=0,{\bm{\delta}}_{-j})}\Big)}^{{\delta}_{j}}\\
	&\quad \times{\Big(\frac{\omega p(\bm{y}|{\delta}_{j}=0,{\bm{\delta}}_{-j})}{\omega p(\bm{y}|{\delta}_{j}=1,{\bm{\delta}}_{-j})+(1-\omega) p(\bm{y}|{\delta}_{j}=0,{\bm{\delta}}_{-j})}\Big)}^{1-{\delta}_{j}},
\end{align*}
where ${\bm{\delta}}_{-j}$ is the vector consisting of all the elements of $\bm{\delta}$ except ${\delta}_{j}$.

In addition, the full conditional posteriors of $(\omega,{\sigma}^{2},{\bm{\beta}}_{\bm{\delta}})$ are given by
\begin{align*}
	\omega|\bm{\delta}&\sim {\rm Beta}({a}_{\omega}+{p}_{1},{b}_{\omega}+p-1-{p}_{1}),\\
	{\sigma}^{2}|\bm{y},\bm{\delta}&\sim {\rm IG} \left(\frac{n}{2},{s}_{c} \right),\\
	{\bm{\beta}}_{\bm{\delta}}|\bm{y},\bm{\delta},{\sigma}^{2}&\sim {\rm N}_{p_1+1}({\bm{h}}_{\bm{\delta}},{\sigma}^{2}{\bm{H}}_{\bm{\delta}}).
\end{align*}
The derivations of the full conditional posteriors are given in Appendix B.

\subsection{Computational algorithm}
\label{sec:ComputAlgo}
With the full conditional posteriors, we can obtain an estimate of the parameters by Gibbs sampling. 
The Gibbs sampling algorithm for our proposed method is as follows:
\begin{description}
	\item[Step 1.] Sample $(\bm{\delta},{\sigma}^{2})$ from the posterior $p(\bm{\delta}|\bm{y})p({\sigma}^{2}|\bm{y},\bm{\delta})$.
	\begin{enumerate}
		\item Sample each element ${\delta}_{j}$ of the indicator vector $\bm{\delta}$ separately from $p({\delta}_{j}=1|{\bm{\delta}}_{-j},\bm{y})$ given as
		\begin{eqnarray*}
		p({\delta}_{j}=1|{\bm{\delta}}_{-j},\bm{y})=\frac{1}{1+\frac{1-\omega}{\omega}{R}_{j}},\quad{R}_{j}=\frac{p(\bm{y}|{\delta}_{j}=0,{\bm{\delta}}_{-j})}{p(\bm{y}|{\delta}_{j}=1,{\bm{\delta}}_{-j})}.
    	\end{eqnarray*}
    		Note that the elements of $\bm{\delta}$ are updated in a random permutation order.
    	\item Sample the error variance ${\sigma}^{2}$ from ${\rm IG}(n/2,{s}_{c})$.
    \end{enumerate}
    \item[Step 2.] Sample $\omega$ from $\omega \sim {\rm Beta}({a}_{\omega}+{p}_{1},{b}_{\omega}+p-1-{p}_{1})$.
    \item[Step 3.] Set ${\beta}_{j}={\beta}_{j+1}$ if ${\delta}_{j}=0$. 
    Sample the other elements ${\bm{\beta}}_{\bm{\delta}}$ from ${\rm N}_{p_1+1}({\bm{h}}_{\bm{\delta}},{\sigma}^{2}{\bm{H}}_{\bm{\delta}})$.
\end{description}

\section{Numerical studies}
\label{sec:NumericalStudies}
In this section, we demonstrate the effectiveness of our proposed method and compare it with previous methods through Monte Carlo simulations. 
In addition, we apply our proposed method to comparative genomic hybridization (CGH) array data.

\subsection{Monte Carlo simulation}
\label{sec:MonteCarlo}
We simulated data with sample size $n$ and number of predictors $p$ from the true model
\begin{eqnarray}
	\bm{y}=\bm{X}{\bm{\beta}}^{*} +\bm{\varepsilon},
\end{eqnarray}
where ${\bm{\beta}}^{*}$ is the $p$-dimensional true coefficient vector, $\bm{X}$ is the $n \times p$ design matrix, and $\bm{\varepsilon}$ is an error vector distributed as ${\rm N}_{n}({\bm{0}}_{n},{\sigma}^{2}{\bm{I}}_{n})$. 
Furthermore, ${\bm{x}}_{i} \ (i=1,\ldots,n)$ was generated from a multivariate normal distribution ${\rm N}_{p}({\bm{0}}_{p},{\bm{\Sigma}})$. 
Let $\Sigma_{ij}$ be the $(i,j)$-th element of ${\bm{\Sigma}}$. 
If $i=j$, then we set $\Sigma_{ij}=1$, and otherwise $\Sigma_{ij}=\rho$. 
We considered $\rho=0,0.5$.
We simulated 100 datasets with different number of observations $n=50, 100, 200$. 
We considered the following six cases: 
\begin{description}
\item[Case 1] ${\bm{\beta}}^{*}={({\bm{1.0}}^{T}_{5},{\bm{1.5}}^{T}_{5},{\bm{1.0}}^{T}_{5},{\bm{1.5}}^{T}_{5})}^{T},\ \sigma =0.75$;
\item[Case 2] ${\bm{\beta}}^{*}={({\bm{1.0}}^{T}_{5},{\bm{1.5}}^{T}_{5},{\bm{1.0}}^{T}_{5},{\bm{1.5}}^{T}_{5})}^{T},\ \sigma =1.5$;
\item[Case 3] ${\bm{\beta}}^{*}={({\bm{1.0}}^{T}_{5},{\bm{2.0}}^{T}_{5},{\bm{1.0}}^{T}_{5},{\bm{2.0}}^{T}_{5})}^{T},\ \sigma =0.75$;
\item[Case 4] ${\bm{\beta}}^{*}={({\bm{1.0}}^{T}_{5},{\bm{2.0}}^{T}_{5},{\bm{1.0}}^{T}_{5},{\bm{2.0}}^{T}_{5})}^{T},\ \sigma =1.5$;
\item[Case 5] ${\bm{\beta}}^{*}={({\bm{1.0}}^{T}_{5},{\bm{3.0}}^{T}_{5},{\bm{1.0}}^{T}_{5},{\bm{3.0}}^{T}_{5})}^{T},\ \sigma =0.75$;
\item[Case 6] ${\bm{\beta}}^{*}={({\bm{1.0}}^{T}_{5},{\bm{3.0}}^{T}_{5},{\bm{1.0}}^{T}_{5},{\bm{3.0}}^{T}_{5})}^{T},\ \sigma =1.5$.
\end{description}

We used $a_\omega=b_\omega=1$, which are included in the prior for the hyper-parameter $\omega$. 
We set the value of the hyper-parameter $g$ in the variance-covariance matrix ${\bm B}_{0,{\bm \delta}}$ to the sample size $n$ because model selection based on Bayes factors is consistent with the $g$-prior with $g=n$ \citep{fernandez2001benchmark}. 
For each dataset, MCMC procedure was run for 10,000 iterations with 2,000 draws as burn-in. 
The parameters were estimated by their posterior means.

\begin{figure}[htbp]
\centering
\includegraphics[scale=1.0]{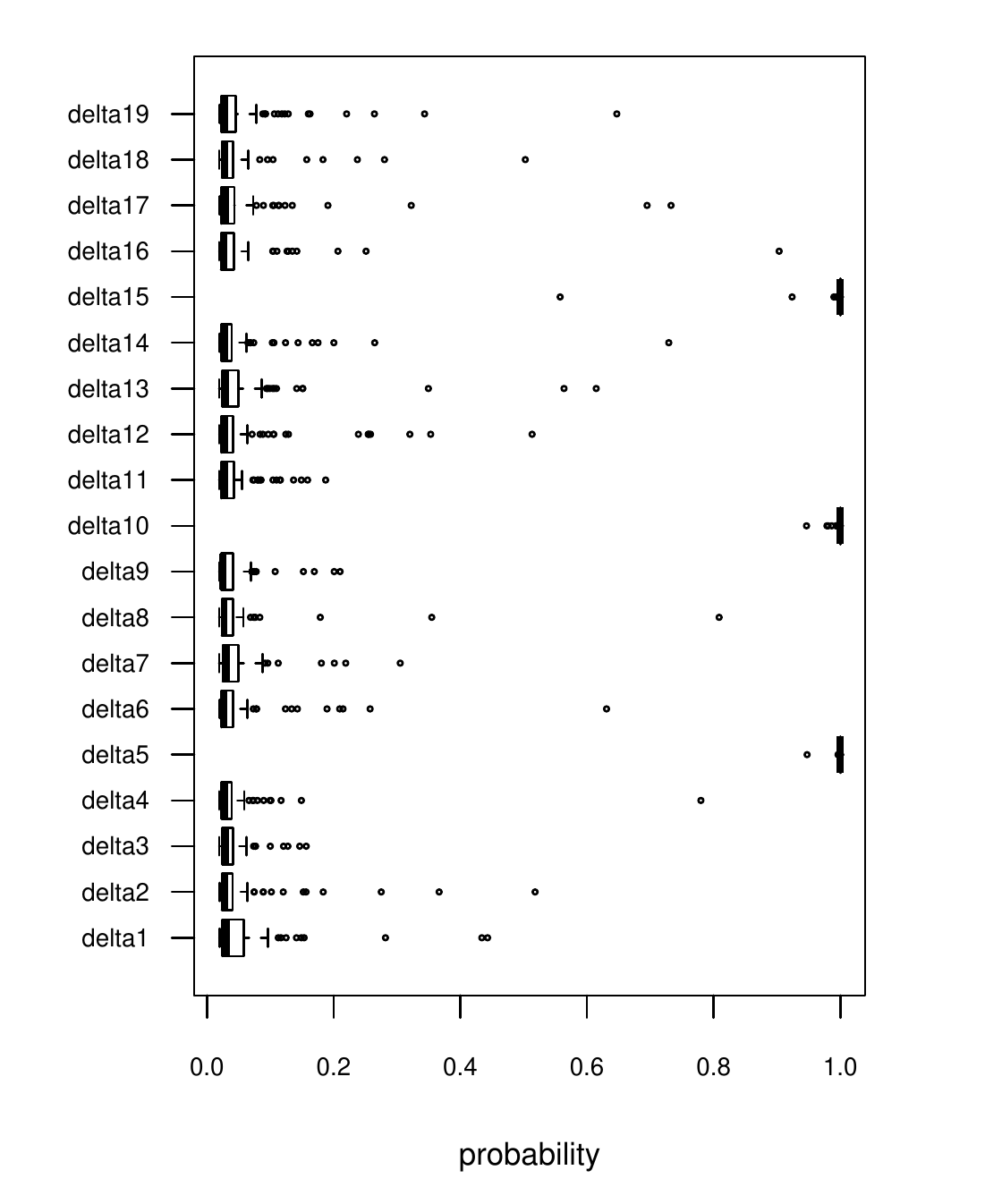}
\caption{Boxplots for estimated probabilities of ${\delta}_{j}=1\ (j=1,\ldots,19)$, from 100 simulations for Case 1 ($n$=200, $\rho=0.5$).}
\label{fig:boxoplot}
\end{figure}

Figure \ref{fig:boxoplot} shows estimated probabilities of ${\delta}_{j}=1\ (j=1,\ldots,19)$ during 100 simulations for Case 1 ($n$=200, $\rho=0.5$). 
We observe that the probabilities of ${\delta}_{1}=1$, ${\delta}_{2}=1$, ${\delta}_{3}=1$, and ${\delta}_{4}=1$ are almost $0\%$, which means ${\beta}_{1}={\beta}_{2}={\beta}_{3}={\beta}_{4}={\beta}_{5}$. 
They are actually in the same group. 
The probability of ${\delta}_{5}=1$ is almost $100\%$, which means ${\beta}_{5}\neq{\beta}_{6}$. 
They are actually not in the same group. 
By the same reasoning, we observe that ${\beta}_{6}$, ${\beta}_{7}$, ${\beta}_{8}$, ${\beta}_{9}$, and ${\beta}_{10}$ seem to be in the same group, and ${\beta}_{11}$, ${\beta}_{12}$, ${\beta}_{13}$, ${\beta}_{14}$, and ${\beta}_{15}$ seem to be in the same group, and ${\beta}_{16}$, ${\beta}_{17}$, ${\beta}_{18}$, ${\beta}_{19}$, and ${\beta}_{20}$ seem to be in the same group. 
This shows that our proposed method can identify the true groups of the coefficients.

We compared our proposed method with the fused lasso (FL), the Bayesian lasso (BFL), and the Bayesian fused lasso via the NEG prior (BFNEG). 
For FL, BFL, and BFNEG, we omitted the lasso penalty term because we only focus on identifying the true groups of coefficients. 
The regularization parameter in the fusion penalty term was selected by the extended Bayesian information criterion (EBIC), which was introduced by \cite{chen2008extended}. 
FL is implemented in the {\tt penalized} package of {\tt R}. 
BFL and BFNEG are implemented in the {\tt neggfl} package of {\tt R}, which is available from \url{https://github.com/ksmstg/neggfl}.

First, we compared the accuracy of prediction using the mean squared error (MSE) and prediction squared error (PSE) given by
\begin{align*}
    {\rm MSE}&=\frac { 1 }{ 100 }  \sum _{  k=1}^{100}{{({\hat{\bm{\beta}}}^{(k)}-{\bm{\beta}}^{*})}^{T}}({\hat{\bm{\beta}}}^{(k)}-{\bm{\beta}}^{*}),\\ 
	{\rm PSE}&=\frac { 1 }{ 100 }  \sum _{  k=1}^{100}{{({\hat{\bm{\beta}}}^{(k)}-{\bm{\beta}}^{*})}^{T}}\Sigma({\hat{\bm{\beta}}}^{(k)}-{\bm{\beta}}^{*}),
\end{align*}
where ${\hat{\bm{\beta}}}^{(k)}$ is the estimated coefficients with the $k$-th dataset and $\Sigma$ is the variance-covariance matrix of the predictors.
Additionally, we compared the accuracy of identifying the true coefficient groups between our proposed method and the competing methods. 
We denote the groups of indexes which have same true regression coefficients by ${B}_{1},\ldots,{B}_{L}\subset\{1,\ldots,p\}$. 
We also denote the number of distinct coefficients in $\{{\hat{\bm{\beta}}}^{(k)}_{j}:j\in {B}_{l}\}\ (l=1,\ldots,L)$ by ${N}_{l}$. 
Finally, we define the accuracy by
\begin{equation*}
	{P}_{B}=\frac{1}{100}\sum^{100}_{k=1}\frac{p-\sum^{L}_{l=1}{N}^{(k)}_{l}}{p-L},
\end{equation*}
so a higher value indicates more accurate variable fusion.

\begin{table}
	\centering
	\tbl{Results for Case 1.}
	{\begin{tabular}{lccccccc}
		\toprule
		 & &\multicolumn{3}{c}{$\rho=0.5$}&\multicolumn{3}{c}{$\rho=0$} \\
		 \cmidrule(r){3-5}\cmidrule(r){6-8}
		 & & MSE &PSE &${P}_{B}$ &MSE&PSE &${P}_{B}$ \\
		 & & (sd)&(sd)& &(sd)&(sd)&\\
		 \hline
		 \multirow{8}{*}{$n$=50}& FL& 0.269&0.146 &0.842 &0.433 &0.433 &0.689 \\
		  &    &(0.154)&(0.079)& &(0.117) &(0.117) &\\
		  & BFL& 0.227&0.128 &0.731 &0.226 &0.226 &0.388\\
		    &    &(0.129)&(0.068)& &(0.091) &(0.091) &\\
		  & BFNEG & 0.535&0.279 &0.336 &0.240 &0.240 &0.369\\
		    &    &(0.251)&(0.127)& &(0.112) &(0.112) &\\
		  & Proposed& 0.210& 0.117& 0.980&0.104 &0.104 &0.971\\
		   &    &(0.174)&(0.089)& &(0.083) &(0.083) &\\
		\bottomrule
		 \multirow{8}{*}{$n$=100}& FL& 0.269&0.146 &0.842 &0.153 &0.153 &0.832 \\
		  &    &(0.154)&(0.079)& &(0.071) &(0.071) &\\
		  & BFL& 0.227&0.128 &0.731 &0.070 &0.070 &0.789\\
		    &    &(0.129)&(0.068)& &(0.039) &(0.039) &\\
		  & BFNEG & 0.535&0.279 &0.336 &0.117 &0.117 &0.321\\
		    &    &(0.251)&(0.127)& &(0.047) &(0.047) &\\
		  & Proposed& 0.210& 0.117& 0.980&0.028 &0.028 &0.991\\
		   &    &(0.174)&(0.089)& &(0.018) &(0.018) &\\		 
		 \bottomrule
		  \multirow{8}{*}{$n$=200}& FL& 0.269&0.146 &0.842 &0.099 &0.099 &0.620\\
		  &    &(0.154)&(0.079)& &(0.018) &(0.018) &\\
		  & BFL& 0.227&0.128 &0.731 &0.035 &0.035 &0.579\\
		    &    &(0.129)&(0.068)& &(0.013) &(0.013) &\\
		  & BFNEG & 0.535&0.279 &0.336 &0.032 &0.032 &0.612\\
		    &    &(0.251)&(0.127)& &(0.017) &(0.017) &\\
		  & Proposed& 0.210& 0.117& 0.980&0.014 &0.014 &0.990\\
		   &    &(0.174)&(0.089)& &(0.009) &(0.009) &\\
		  \bottomrule
	\end{tabular}}
	\label{tab:MonteResult1}
\end{table}

\begin{table}
	\centering
	\tbl{Results for Case 2.}
	{\begin{tabular}{lccccccc}
		\toprule
		 & &\multicolumn{3}{c}{$\rho=0.5$}&\multicolumn{3}{c}{$\rho=0$} \\
		 \cmidrule(r){3-5}\cmidrule(r){6-8}
		 & & MSE &PSE &${P}_{B}$ &MSE&PSE &${P}_{B}$ \\
		 & & (sd)&(sd)& &(sd)&(sd)&\\
		 \hline
		 \multirow{8}{*}{$n$=50}& FL& 0.946&0.529 &0.933 &0.775 &0.775 &0.862 \\
		  &    &(0.217)&(0.127)& &(0.642) &(0.642) &\\
		  & BFL& 2.987&1.550 &0.134 &0.532 &0.532 &0.626\\
		    &    &(1.334)&(0.672)& &(0.262) &(0.262) &\\
		  & BFNEG & 1.508&0.792 &0.657 &0.800 &0.800 &0.442\\
		    &    &(0.842)&(0.422)& &(0.376) &(0.376) &\\
		  & Proposed& 0.782& 0.430& 0.987&0.584 &0.584 &0.977\\
		   &    &(0.280)&(0.163)& &(0.258) &(0.258) &\\
		\bottomrule
		 \multirow{8}{*}{$n$=100}& FL& 0.386&0.217 &0.761 &0.247 &0.247 &0.803 \\
		  &    &(0.192)&(0.106)& &(0.142) &(0.142) &\\
		  & BFL& 0.467&0.263 &0.869 &0.428 &0.428 &0.375\\
		    &    &(0.234)&(0.127)& &(0.194) &(0.194) &\\
		  & BFNEG & 0.942&0.496 &0.210 &0.336 &0.336 &0.806\\
		    &    &(0.359)&(0.179)& &(0.136) &(0.136) &\\
		  & Proposed& 0.485& 0.265& 0.984&0.228 &0.228 &0.971\\
		   &    &(0.253)&(0.132)& &(0.176) &(0.176) &\\		 
		 \bottomrule
		  \multirow{8}{*}{$n$=200}& FL& 0.202&0.113 &0.755 &0.108 &0.108 &0.654\\
		  &    &(0.122)&(0.063)& &(0.053) &(0.053) &\\
		  & BFL& 0.257&0.137 &0.848 &0.155 &0.155 &0.560\\
		    &    &(0.133)&(0.068)& &(0.067) &(0.067) &\\
		  & BFNEG & 0.339&0.182 &0.625 &0.221 &0.221 &0.921\\
		    &    &(0.162)&(0.084)& &(0.088) &(0.088) &\\
		  & Proposed& 0.188& 0.103& 0.989&0.066 &0.066 &0.998\\
		   &    &(0.163)&(0.083)& &(0.049) &(0.049) &\\
		  \bottomrule
	\end{tabular}}
	\label{tab:MonteResult2}
\end{table}

\begin{table}
	\tbl{Results for Case 3.}
	{\begin{tabular}{lccccccc}
		\toprule
		 & &\multicolumn{3}{c}{$\rho=0.5$}&\multicolumn{3}{c}{$\rho=0$} \\
		 \cmidrule(r){3-5}\cmidrule(r){6-8}
		 & & MSE &PSE &${P}_{B}$ &MSE&PSE &${P}_{B}$ \\
		 & & (sd)&(sd)& &(sd)&(sd)&\\
		 \hline
		 \multirow{8}{*}{$n$=50}& FL& 0.245&0.136 &0.671 &0.395 &0.395 &0.822 \\
		  &    &(0.165)&(0.084)& &(0.171) &(0.171) &\\
		  & BFL& 0.286&0.155 &0.677 &0.243 &0.243 &0.393\\
		    &    &(0.146)&(0.076)& &(0.111) &(0.111) &\\
		  & BFNEG & 0.462&0.247&0.549 &0.308 &0.308 &0.238\\
		    &    &(0.211)&(0.112)& &(0.130) &(0.130) &\\
		  & Proposed& 0.127& 0.078& 0.978&0.068 &0.068 &0.997\\
		   &    &(0.115)&(0.061)& &(0.047) &(0.047) &\\
		\bottomrule
		 \multirow{8}{*}{$n$=100}& FL& 0.117&0.065 &0.790 &0.093 &0.093 &0.701 \\
		  &    &(0.060)&(0.031)& &(0.032) &(0.032) &\\
		  & BFL& 0.178&0.096 &0.354 &0.061 &0.061 &0.763\\
		    &    &(0.070)&(0.038)& &(0.027) &(0.027) &\\
		  & BFNEG & 0.180&0.097 &0.669 &0.068 &0.068 &0.425\\
		    &    &(0.084)&(0.084)& &(0.029) &(0.029) &\\
		  & Proposed& 0.042& 0.027& 1.000&0.027 &0.027 &0.995\\
		   &    &(0.032)&(0.019)& &(0.019) &(0.019) &\\		 
		 \bottomrule
		  \multirow{8}{*}{$n$=200}& FL& 0.051&0.028 &0.741 &0.032 &0.032 &0.736\\
		  &    &(0.032)&(0.016)& &(0.028) &(0.028) &\\
		  & BFL& 0.092&0.048 &0.307 &0.032 &0.032 &0.678\\
		    &    &(0.035)&(0.018)& &(0.015) &(0.015) &\\
		  & BFNEG & 0.095&0.050 &0.535 &0.050 &0.050 &0.351\\
		    &    &(0.045)&(0.023)& &(0.017) &(0.017) &\\
		  & Proposed& 0.021& 0.013& 1.000&0.014 &0.014 &0.999\\
		   &    &(0.019)&(0.011)& &(0.010) &(0.010) &\\
		  \bottomrule
	\end{tabular}}
	\label{tab:MonteResult3}
\end{table}

\begin{table}
    \tbl{Results for Case 4.}
	{\begin{tabular}{lccccccc}
		\toprule
		 & &\multicolumn{3}{c}{$\rho=0.5$}&\multicolumn{3}{c}{$\rho=0$} \\
		 \cmidrule(r){3-5}\cmidrule(r){6-8}
		 & & MSE &PSE &${P}_{B}$ &MSE&PSE &${P}_{B}$ \\
		 & & (sd)&(sd)& &(sd)&(sd)&\\
		 \hline
		 \multirow{8}{*}{$n$=50}& FL& 0.913&0.506 &0.814 &0.815 &0.815 &0.546 \\
		  &    &(0.576)&(0.301)& &(0.306) &(0.306) &\\
		  & BFL& 1.011&0.559 &0.823 &0.755 &0.755 &0.525\\
		    &    &(0.554)&(0.292)& &(0.394) &(0.394) &\\
		  & BFNEG & 1.443&0.786 &0.687 &1.189 &1.189 &0.149\\
		    &    &(0.698)&(0.375)& &(0.479) &(0.479) &\\
		  & Proposed& 1.033& 0.559& 0.959&0.408 &0.408 &0.966\\
		   &    &(0.745)&(0.379)& &(0.380) &(0.380) &\\
		\bottomrule
		 \multirow{8}{*}{$n$=100}& FL& 0.446&0.248 &0.593 &0.319 &0.319 &0.809 \\
		  &    &(0.273)&(0.140)& &(0.157) &(0.157) &\\
		  & BFL& 0.474&0.259 &0.762 &0.324 &0.324 &0.506\\
		    &    &(0.238)&(0.124)& &(0.145) &(0.145) &\\
		  & BFNEG & 0.778&0.410 &0.443 &0.395 &0.395 &0.391\\
		    &    &(0.284)&(0.146)& &(0.156) &(0.156) &\\
		  & Proposed& 0.248& 0.144& 0.975&0.130 &0.130 &0.992\\
		   &    &(0.237)&(0.119)& &(0.082) &(0.082) &\\		 
		 \bottomrule
		  \multirow{8}{*}{$n$=200}& FL& 0.185&0.102 &0.723 &0.138 &0.138 &0.842\\
		  &    &(0.100)&(0.052)& &(0.073) &(0.073) &\\
		  & BFL& 0.255&0.138 &0.854 &0.127 &0.127 &0.877\\
		    &    &(0.131)&(0.068)& &(0.069) &(0.069) &\\
		  & BFNEG & 0.294&0.160 &0.719 &0.162 &0.162 &0.222\\
		    &    &(0.154)&(0.079)& &(0.061) &(0.061) &\\
		  & Proposed& 0.081& 0.053& 0.999&0.054 &0.054 &0.996\\
		   &    &(0.058)&(0.033)& &(0.035) &(0.035) &\\
		  \bottomrule
	\end{tabular}}
	\label{tab:MonteResult4}
\end{table}

\begin{table}
    \tbl{Results for Case 5.}
	{\begin{tabular}{lccccccc}
		\toprule
		 & &\multicolumn{3}{c}{$\rho=0.5$}&\multicolumn{3}{c}{$\rho=0$} \\
		 \cmidrule(r){3-5}\cmidrule(r){6-8}
		 & & MSE &PSE &${P}_{B}$ &MSE&PSE &${P}_{B}$ \\
		 & & (sd)&(sd)& &(sd)&(sd)&\\
		 \hline
		 \multirow{8}{*}{$n$=50}& FL& 0.244&0.134 &0.812 &0.675 &0.675 &0.865 \\
		  &    &(0.217)&(0.109)& &(0.182) &(0.182) &\\
		  & BFL& 0.312&0.168 &0.694 &0.321 &0.321 &0.231\\
		    &    &(0.179)&(0.089)& &(0.141) &(0.141) &\\
		  & BFNEG & 0.497&0.264 &0.549 &0.327 &0.327 &0.212\\
		    &    &(0.229)&(0.121)& &(0.135) &(0.135) &\\
		  & Proposed& 0.103& 0.061& 0.988&0.065 &0.065 &0.987\\
		   &    &(0.079)&(0.041)& &(0.045) &(0.045) &\\
		\bottomrule
		 \multirow{8}{*}{$n$=100}& FL& 0.141&0.077 &0.851 &0.289 &0.289 &0.769 \\
		  &    &(0.087)&(0.046)& &(0.051) &(0.051) &\\
		  & BFL& 0.219&0.116 &0.243 &0.076 &0.076 &0.607\\
		    &    &(0.085)&(0.043)& &(0.034) &(0.034) &\\
		  & BFNEG & 0.202&0.110 &0.602 &0.111 &0.111 &0.295\\
		    &    &(0.096)&(0.048)& &(0.043) &(0.043) &\\
		  & Proposed& 0.043& 0.028& 1.000&0.025 &0.025 &0.993\\
		   &    &(0.032)&(0.018)& &(0.017) &(0.017) &\\		 
		 \bottomrule
		  \multirow{8}{*}{$n$=200}& FL& 0.053&0.029 &0.775 &0.048 &0.048 &0.895\\
		  &    &(0.030)&(0.015)& &(0.024) &(0.024) &\\
		  & BFL& 0.122&0.065 &0.158 &0.051 &0.051 &0.364\\
		    &    &(0.043)&(0.023)& &(0.018) &(0.018) &\\
		  & BFNEG & 0.096&0.051 &0.583 &0.056 &0.056 &0.176\\
		    &    &(0.041)&(0.021)& &(0.019) &(0.019) &\\
		  & Proposed& 0.025& 0.014& 1.000&0.014 &0.014 &0.994\\
		   &    &(0.023)&(0.013)& &(0.009) &(0.009) &\\
		  \bottomrule
	\end{tabular}}
	\label{tab:MonteResult5}
\end{table}

\begin{table}
    \tbl{Results for Case 6.}
	{\begin{tabular}{lccccccc}
		\toprule
		 & &\multicolumn{3}{c}{$\rho=0.5$}&\multicolumn{3}{c}{$\rho=0$} \\
		 \cmidrule(r){3-5}\cmidrule(r){6-8}
		 & & MSE &PSE &${P}_{B}$ &MSE&PSE &${P}_{B}$ \\
		 & & (sd)&(sd)& &(sd)&(sd)&\\
		 \hline
		 \multirow{8}{*}{$n$=50}& FL& 1.093&0.593 &0.786 &1.419 &1.419&0.833 \\
		  &    &(0.834)&(0.426)& &(0.412) &(0.412) &\\
		  & BFL& 1.241&0.660 &0.634 &1.026 &1.026 &0.333\\
		    &    &(0.665)&(0.334)& &(0.431) &(0.431) &\\
		  & BFNEG & 1.618&0.863 &0.582 &0.869 &0.869 &0.508\\
		    &    &(0.744)&(0.387)& &(0.396) &(0.396) &\\
		  & Proposed& 0.501& 0.305& 0.993&0.266 &0.266 &0.985\\
		   &    &(0.418)&(0.236)& &(0.178) &(0.178) &\\
		\bottomrule
		 \multirow{8}{*}{$n$=100}& FL& 0.383&0.213 &0.653 &0.677 &0.677 &0.780 \\
		  &    &(0.211)&(0.108)& &(0.160) &(0.160) &\\
		  & BFL& 0.515&0.284 &0.679 &0.280 &0.280 &0.834\\
		    &    &(0.265)&(0.138)& &(0.162) &(0.162) &\\
		  & BFNEG & 0.642&0.354 &0.667 &0.496 &0.496 &0.206\\
		    &    &(0.295)&(0.163)& &(0.179) &(0.179) &\\
		  & Proposed& 0.174& 0.103& 0.994&0.134 &0.134 &0.990\\
		   &    &(0.146)&(0.082)& &(0.096) &(0.096) &\\		 
		 \bottomrule
		  \multirow{8}{*}{$n$=200}& FL& 0.202&0.110 &0.673 &0.248 &0.248 &0.876\\
		  &    &(0.104)&(0.052)& &(0.073) &(0.073) &\\
		  & BFL& 0.222&0.123 &0.743 &0.143 &0.143 &0.583\\
		    &    &(0.108)&(0.053)& &(0.057) &(0.057) &\\
		  & BFNEG & 0.345&0.183 &0.590 &0.220 &0.220 &0.197\\
		    &    &(0.145)&(0.073)& &(0.069) &(0.069) &\\
		  & Proposed& 0.093& 0.056& 1.000&0.055 &0.055 &0.992\\
		   &    &(0.075)&(0.039)& &(0.039) &(0.039) &\\
		  \bottomrule
	\end{tabular}}
		\label{tab:MonteResult6}
\end{table}

The simulation results are summarized in Tables \ref{tab:MonteResult1} to \ref{tab:MonteResult6}. 
First, in almost all cases, all methods showed higher MSEs and PSEs in the estimation for $\rho=0.5$ than those for $\rho=0$. 
FL showed higher values of ${P}_{B}$ than BFL and BFNEG in almost all cases. 
Its average accuracy was about 0.75. 
Both BFL and BFNEG showed unstable values of ${P}_{B}$ regardless of the setup, which means that they cannot provide stability for identifying the true groups.
Almost all criteria indicated that our proposed method provides much better performance than the competing methods. 
In particular, the values of ${P}_{B}$ were close to one in all cases. 
This means that the true groups are identified almost perfectly by our proposed method. 
Furthermore, the low MSEs and PSEs showed that our proposed method cannot only identify true groups but also provides good estimates for the true regression coefficients.


\subsection{Application}
\label{sec:RealData}
We applied our proposed method to a smoothing illustration for a real dataset. 
We used the comparative genomic hybridization (CGH) array data \citep{tibshirani2007spatial}. 
The data are available in the {\tt cghFLasso} package of the software {\tt R}.
The data comprise log ratios of the genome numbers of copies between a normal cell and a cancer cell by genome order. 
We extracted 150 samples from the genome orders 50 to 200 and set them as $\bm{y}$. 
The design matrix $\bm{X}$ was set as an identity matrix. 
Thus, the model used in this section was $\bm{y}=\bm{\beta}+\bm{\varepsilon}$. 
We compared our proposed method with FL, BFL, and BFNEG. 
We determined the values of the hyper-parameters in our proposed method in a similar manner to that in Section \ref{sec:MonteCarlo}. 
The values of the hyper-parameters in the other methods were selected by the EBIC.

\begin{figure}[htbp]
\centering
\includegraphics[scale=0.57]{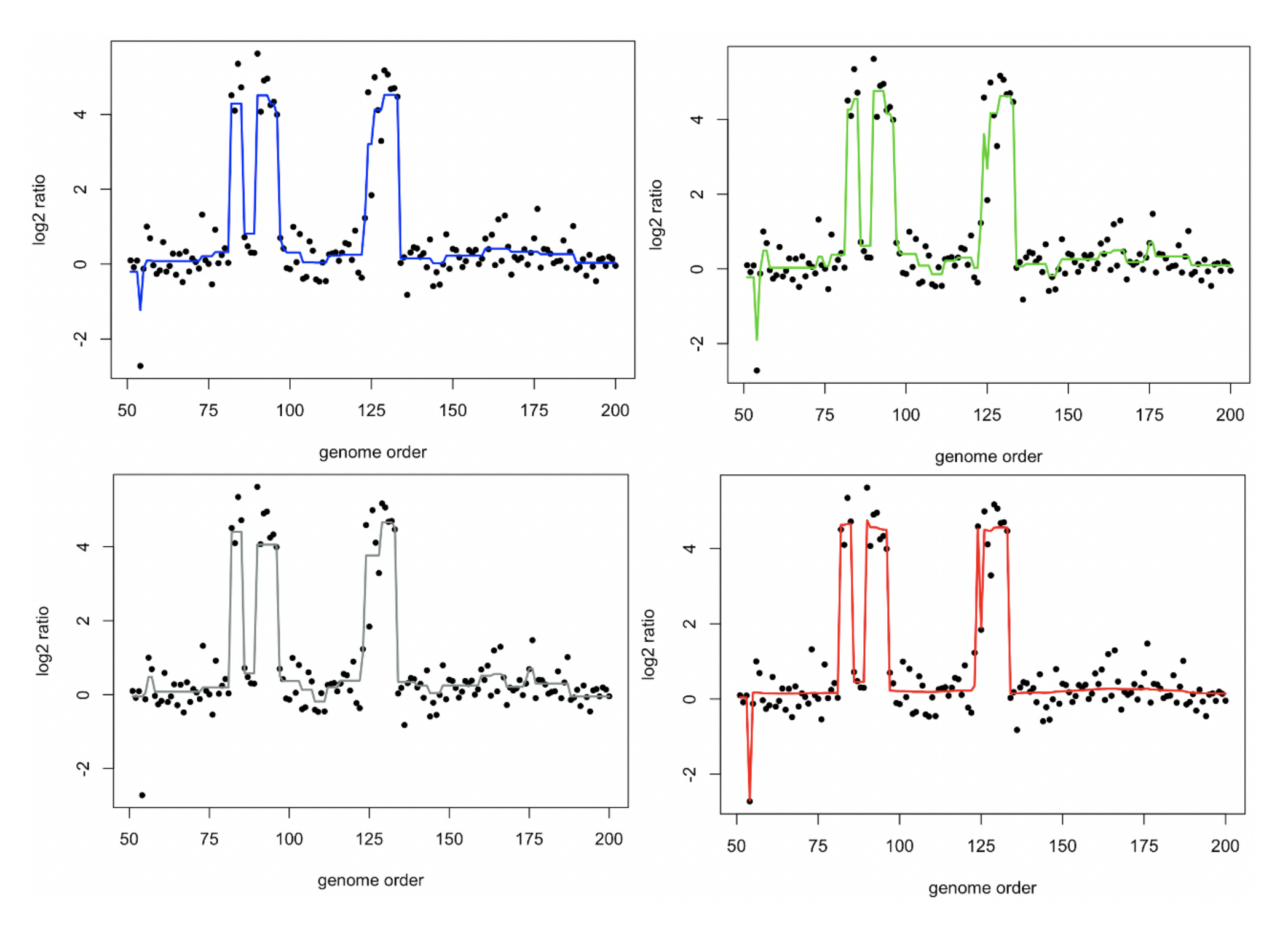}
\caption{Results for the CGH data. 
Black dots indicate data points.
The blue line is the estimates of FL, the green line is  BFL, the grey line is  BFNEG, and the red line is  our proposed method.}
\label{fig:CGH}
\end{figure}

Figure \ref{fig:CGH} shows the result for applying the methods to the CGH array data. 
The data points from about 55 to 80, 95 to 120, and 135 to 200 showed that the genome copy numbers of cancer cells and normal cells are almost the same. 
We can see that our proposed method gives stable estimates, while the competing methods seem to overfit. 
The other data points showed that the genome copy numbers of cancer cells are bigger than those of the normal cells. 
Our proposed method captures differences more clearly than the competing methods.

\section{Conclusions}
\label{sec:Conclusions}
In this work, we have discussed the Bayesian variable fusion method from the viewpoint of linear regression models. 
We have proposed a spike-and-slab prior that induces variable fusion, which is based on the Dirac spike and the $g$-slab prior. 
To obtain samples from the posteriors, a Gibbs sampler was designed using the hierarchical Bayesian models from the spike-and-slab prior. 
Numerical studies showed that our proposed method performs well compared to existing methods from the various viewpoints.

Our proposed method cannot perform variable selection, unlike previous Bayesian methods \citep{kyung2010penalized,shimamura2019bayesian}. 
Therefore, it would be interesting to extend our proposed method to handle variable selection. 
We leave this topic as future research.

\section*{Acknowledgments}
SK was supported by JSPS KAKENHI Grant Number JP19K11854 and JP20H02227 and MEXT KAKENHI Grant Numbers JP16H06429, JP16K21723, and JP16H06430.

\appendix

\section{Marginal likelihood}

First, the marginal likelihood $p(\bm{y}|\bm{\delta})$ can be denoted by $p(\bm{y}|{\bm{X}}_{\bm{\delta}})$, because $\bm{\delta}$ has a one-to-one relationship with the design matrix ${\bm{X}}_{\bm{\delta}}$. 
We then consider the likelihood of the linear regression model with ${\bm{X}}_{\bm{\delta}}$ and ${\bm{\beta}}_{\bm{\delta}}$ in the form
\begin{equation*}
	 p(\bm{y}|{\bm{X}}_{\bm{\delta}},{\bm{\beta}}_{\bm{\delta}} ,{ \sigma  }^{ 2 })=\frac { 1 }{ { (2\pi ) }^{ n/2 }{ |{ \sigma  }^{ 2 }\bm{I}_n| }^{ 1/2 } } \exp \left\{ -\frac { 1 }{ 2 } { (\bm{y}-{\bm{X}}_{\bm{\delta}}{\bm{\beta}}_{\bm{\delta}} ) }^{ T }{ ({ \sigma  }^{ 2 }\bm{I}_n) }^{ -1 }(\bm{y}-{\bm{X}}_{\bm{\delta}}{\bm{\beta}}_{\bm{\delta}} ) \right\}.
\end{equation*}
The joint distribution $p(\bm{y}|{\bm{X}}_{\bm{\delta}},{\bm{\beta}}_{\bm{\delta}} ,{ \sigma  }^{ 2 })p({\bm{D}}_{\bm{\beta}}{\bm{\beta}}_{\bm{\delta}} )p({\sigma}^{2})$ is calculated as
\begin{align*}
	 &p(\bm{y}|{\bm{X}}_{\bm{\delta}},{\bm{\beta}}_{\bm{\delta}} ,{ \sigma  }^{ 2 })p({\bm{D}}_{\bm{\beta}}{\bm{\beta}}_{\bm{\delta}} )p({\sigma}^{2})\\
	 &=\frac { 1 }{ { (2\pi ) }^{ n/2 }{ |{ \sigma  }^{ 2 }\bm{I}_n| }^{ 1/2 } }\frac { 1 }{ { (2\pi ) }^{ {p}_{1}/2 }{ |{\sigma}^{2}{ \bm{H} }_{ 0 ,\bm{\delta}}| }^{ 1/2 } } \\ 
	 &\quad\times \exp \left\{ -\frac { 1 }{ 2 } { (\bm{y}-{\bm{X}}_{\bm{\delta}}{\bm{\beta}}_{\bm{\delta}} ) }^{ T }{ ({ \sigma  }^{ 2 }\bm{I}_n) }^{ -1 }(\bm{y}-{\bm{X}}_{\bm{\delta}}{\bm{\beta}}_{\bm{\delta}} ) \right\}\\
     &\quad\times \exp \left\{ -\frac { 1 }{ 2 } { ({\bm{D}}_{\bm{\beta}}{\bm{\beta}}_{\bm{\delta}}-{ \bm{h} }_{ 0 ,\bm{\delta}}) }^{ T }{ ({\sigma}^{2}{ \bm{H} }_{ 0 ,\bm{\delta}}) }^{ -1 }({\bm{D}}_{\bm{\beta}}{\bm{\beta}}_{\bm{\delta}} -{ \bm{h} }_{ 0,\bm{\delta} }) \right\}\\
     &\quad\times \frac{1}{{\sigma}^{2}}\\
     &= \frac { 1 }{ { (2\pi ) }^{ n/2 } {{ \sigma  }^{ 2 }}^{(n+2)/2}}\frac { 1 }{ { (2\pi ) }^{ {p}_{1}/2 }{ |{\sigma}^{2}{ \bm{H} }_{ 0 ,\bm{\delta}}| }^{ 1/2 } } \\
     &\quad\times\exp \bigg\{ -\frac { 1 }{ 2 } [{ {\bm{\beta}}_{\bm{\delta}}  }^{ T }({ \bm{X} }^{ T }_{\bm{\delta}}{ ({ \sigma  }^{ 2 }\bm{I}_n) }^{ -1 }{\bm{X}}_{\bm{\delta}}+{ \bm{D} }^{ T }_{\bm{\delta}}{ ({ \sigma  }^{ 2 }{ \bm{H} }_{ 0 ,\bm{\delta}}) }^{ -1 }{\bm{D}}_{\bm{\delta}}){\bm{\beta}}_{\bm{\delta}} -2{ \bm{\beta}  }^{ T }_{\bm{\delta}}({ \bm{X} }^{ T }_{\bm{\delta}}{ ({ \sigma  }^{ 2 }\bm{I}_n) }^{ -1 }\bm{y}\\
     &\quad +{ \bm{D} }^{ T }_{\bm{\delta}}{ ({ \sigma  }^{ 2 }{ \bm{H} }_{ 0 ,_{\bm{\delta}}}) }^{ -1 }{ \bm{h} }_{ 0 ,\bm{\delta}})+{ \bm{y} }^{ T }{ ({ \sigma  }^{ 2 }\bm{I}_n) }^{ -1 }\bm{y}+{ \bm{h} }_{ 0,\bm{\delta} }^{ T }{ ({ \sigma  }^{ 2 }{ \bm{H} }_{ 0 ,\bm{\delta}}) }^{ -1 }{ \bm{h} }_{ 0,\bm{\delta} }] \bigg\}\\
     &=\frac { 1 }{ { (2\pi ) }^{ n/2 } {{ \sigma  }^{ 2 }}^{(n+2)/2}}\frac { 1 }{ { (2\pi ) }^{ {p}_{1}/2 }{ |{\sigma}^{2}{ \bm{H} }_{ 0 ,\bm{\delta}}| }^{ 1/2 } } \\
     &\quad\times\exp \bigg\{-\frac { 1 }{ 2{ \sigma  }^{ 2 } } [{ \bm{\beta}  }^{ T }_{\bm{\delta}}({ \bm{X} }^{ T }_{\bm{\delta}}{\bm{X}}_{\bm{\delta}}+{ \bm{D} }^{ T }_{\bm{\delta}}{ { \bm{H} }_{ 0 ,\bm{\delta}} }^{ -1 }{\bm{D}}_{\bm{\delta}}){\bm{\beta}}_{\bm{\delta}} -2{ \bm{\beta}  }^{ T }_{\bm{\delta}}({ \bm{X} }^{ T }_{\bm{\delta}}\bm{y}+{ \bm{D} }^{ T }_{\bm{\delta}}{ { \bm{H} }_{ 0 ,\bm{\delta}}^{ -1 } }{ \bm{h} }_{ 0 ,\bm{\delta}})\\
     &\quad+{ \bm{y} }^{ T }\bm{y}+{ \bm{h} }_{ 0,\bm{\delta} }^{ T }{ { \bm{H} }_{ 0,\bm{\delta} }^{ -1 } }{ \bm{h} }_{ 0 ,\bm{\delta}}] \bigg\}. 
\end{align*}
By defining
\begin{equation*}
	{\bm{H}}_{\bm{\delta}}={ ({ \bm{X} }^{ T }_{\bm{\delta}}{\bm{X}}_{\bm{\delta}}+{ {\bm{D}}_{\bm{\delta}} }^{ T }{ \bm{H} }_{ 0 ,\bm{\delta}}^{ -1 }{\bm{D}}_{\bm{\delta}}) }^{ -1 },\quad 
	{ \bm{h} }_{ \bm{\delta} }={ \bm{H}}_{ \bm{\delta}  }({ \bm{X} }^{ T }_{\bm{\delta}}\bm{y}+{ \bm{D} }^{ T }_{\bm{\delta}}{ \bm{H} }_{ 0 ,\bm{\delta}}^{ -1 }{ \bm{h} }_{ 0 ,\bm{\delta}}),
\end{equation*}
we can obtain
\begin{align*}
	&p(\bm{y}|{\bm{X}}_{\bm{\delta}},{\bm{\beta}}_{\bm{\delta}} ,{ \sigma  }^{ 2 })p({\bm{D}}_{\bm{\beta}}{\bm{\beta}}_{\bm{\delta}} )p({\sigma}^{2})\\
    &=\frac { 1 }{ { (2\pi ) }^{ n/2 } {{ \sigma  }^{ 2 }}^{(n+2)/2}}\frac { 1 }{ { (2\pi ) }^{ {p}_{1}/2 }{ |{\sigma}^{2}{ \bm{H} }_{ 0 ,\bm{\delta}}| }^{ 1/2 } } \\
    &\quad\times\exp \left\{ -\frac { 1 }{ 2{ \sigma  }^{ 2 } } [{ \bm{\beta}  }^{ T }_{\bm{\delta}}{ \bm{H} }_{ \bm{\delta}  }^{ -1 }{\bm{\beta}}_{\bm{\delta}} +2{ \bm{\beta}  }^{ T }_{\bm{\delta}}{ \bm{H} }_{ \bm{\delta}  }^{ -1 }{ \bm{h} }_{ \bm{\delta}  }+{ \bm{y} }^{ T }\bm{y}+{ \bm{h} }_{ 0,\bm{\delta} }^{ T }{ \bm{H} }_{ 0, \bm{\delta}  }^{ -1 }{ \bm{h} }_{ 0 ,\bm{\delta}}] \right\} \\
    &=\frac { 1 }{ { (2\pi ) }^{ n/2 } {{ \sigma  }^{ 2 }}^{(n+2)/2}}\frac { 1 }{ { (2\pi ) }^{ {p}_{1}/2 }{ |{\sigma}^{2}{ \bm{H} }_{ 0 ,\bm{\delta}}| }^{ 1/2 } } \\
    &\quad\times\exp \left\{ -\frac { 1 }{ 2{ \sigma  }^{ 2 } } ({ \bm{\beta}  }^{ T }_{\bm{\delta}}{ \bm{H} }_{ \bm{\delta}  }^{ -1 }{\bm{\beta}}_{\bm{\delta}} +2{ \bm{\beta}  }^{ T }_{\bm{\delta}}{ \bm{H} }_{ \bm{\delta}  }^{ -1 }{ \bm{h} }_{ \bm{\delta}  }+{ \bm{h} }_{ \bm{\delta}  }^{ T }{ \bm{H} }_{ \bm{\delta}  }^{ -1 }{ \bm{h} }_{ \bm{\delta}  }) \right\}\\
	&\quad \times \exp \left\{ -\frac { 1 }{ 2{ \sigma  }^{ 2 } } ({ \bm{y} }^{ T }\bm{y}+{ \bm{h} }_{ 0,\bm{\delta} }^{ T }{ \bm{H} }_{ 0,\bm{\delta} }^{ -1 }{ \bm{h} }_{ 0,\bm{\delta} }-{ \bm{h} }_{ \bm{\delta}  }^{ T }{ \bm{H} }_{ \bm{\delta}  }^{ -1 }{ \bm{h} }_{ \bm{\delta}  }) \right\} \\
	&=\frac { 1 }{ { (2\pi ) }^{ n/2 } {{ \sigma  }^{ 2 }}^{(n+2)/2}}\frac { 1 }{ { (2\pi ) }^{ {p}_{1}/2 }{ |{\sigma}^{2}{ \bm{H} }_{ 0 ,\bm{\delta}}| }^{ 1/2 } } \\
	&\quad\times\exp \left\{ -\frac { 1 }{ 2 } { ({\bm{\beta}}_{\bm{\delta}} -{ \bm{h} }_{ \bm{\delta}  }) }^{ T }{ ({ \sigma  }^{ 2 }{ \bm{H} }_{ \bm{\delta}  }) }^{ -1 }({\bm{\beta}}_{\bm{\delta}} -{ \bm{h} }_{ \bm{\delta}  }) \right\} \exp \left(-\frac { { s }_{ c } }{ { \sigma  }^{ 2 } } \right),
\end{align*}
where ${s}_{c}=\frac { 1 }{ 2 } ({ \bm{y} }^{ T }\bm{y}+{ \bm{h} }_{ 0 ,\bm{\delta}}^{ T }{ \bm{H} }_{ 0 ,\bm{\delta}}^{ -1 }{ \bm{h} }_{ 0 ,\bm{\delta}}-{ \bm{h} }_{ \bm{\delta}  }^{ T }{ \bm{H} }_{ \bm{\delta}  }^{ -1 }{ \bm{h} }_{ \bm{\delta}  })$.

To derive the marginal likelihood $p(\bm{y}|{\bm{X}}_{\bm{\delta}})$, we first integrate over ${\bm{\beta}}_{\bm{\delta}}$:
\begin{align*}
	p(\bm{y}|{\bm{X}}_{\bm{\delta}} ,{ \sigma  }^{ 2 })p({\sigma}^{2})&=\int p(\bm{y}|{\bm{X}}_{\bm{\delta}},{\bm{\beta}}_{\bm{\delta}} ,{ \sigma  }^{ 2 })p({\bm{D}}_{\bm{\beta}}{\bm{\beta}}_{\bm{\delta}} )p({\sigma}^{2})d{\bm{\beta}}_{\bm{\delta}}\\
	&=\int \frac { {(2\pi)}^{1/2} }{ { (2\pi ) }^{ n/2 } {{ \sigma  }^{ 2 }}^{(n+2)/2}}\frac{ {|{\sigma}^{2}{ \bm{H} }_{ \bm{\delta} }|}^{1/2}}{ {|{\sigma}^{2}{ \bm{H} }_{ 0,\bm{\delta} }|}^{1/2}}\frac { 1 }{ { (2\pi ) }^{ ({p}_{1}+1)/2 }{ |{\sigma}^{2}{ \bm{H} }_{ \bm{\delta} }| }^{ 1/2 } } \\
	 &\quad \times \exp \left\{ -\frac { 1 }{ 2 } { ({\bm{\beta}}_{\bm{\delta}} -{ \bm{h} }_{ \bm{\delta}  }) }^{ T }{ ({ \sigma  }^{ 2 }{ \bm{H} }_{ \bm{\delta}  }) }^{ -1 }({\bm{\beta}}_{\bm{\delta}} -{ \bm{h} }_{ \bm{\delta}  }) \right\} \exp \left(-\frac { { s }_{ c } }{ { \sigma  }^{ 2 } } \right)d{\bm{\beta}}_{\bm{\delta}}\\
	 &= \frac { 1 }{ { (2\pi ) }^{ (n-1)/2 } {{ \sigma  }^{ 2 }}^{(n+2)/2}}\frac{ {|{ \bm{H} }_{ \bm{\delta} }|}^{1/2}}{ {|{ \bm{H} }_{ 0,\bm{\delta} }|}^{1/2}}\exp \left(-\frac { { s }_{ c } }{ { \sigma  }^{ 2 } } \right).
\end{align*}
Next we integrate over ${\sigma}^{2}$:
\begin{align*}
	p(\bm{y}|{\bm{X}}_{\bm{\delta}})&=\int p(\bm{y}|{\bm{X}}_{\bm{\delta}} ,{ \sigma  }^{ 2 })p({\sigma}^{2})d{ \sigma  }^{ 2 }\\
	&=\int \frac { 1 }{ { (2\pi ) }^{ (n-1)/2 } {{ \sigma  }^{ 2 }}^{(n+2)/2}}\frac{ {|{ \bm{H} }_{ \bm{\delta} }|}^{1/2}}{ {|{ \bm{H} }_{ 0,\bm{\delta} }|}^{1/2}}\exp \left(-\frac { { s }_{ c } }{ { \sigma  }^{ 2 } } \right)d{\sigma}^{2}\\
	&=\frac { 1 }{ { (2\pi ) }^{ (n-1)/2 } }\frac{ {|{ \bm{H} }_{ \bm{\delta} }|}^{1/2}}{ {|{ \bm{H} }_{ 0,\bm{\delta} }|}^{1/2}}\int {({\sigma}^{2})}^{-(n+2)/2}\exp \left(-\frac { { s }_{ c } }{ { \sigma  }^{ 2 } } \right)d{\sigma}^{2}\\
	&=\frac { 1 }{ { (2\pi ) }^{ (n-1)/2 } } \frac { { |{ \bm{H} }_{ { \delta  } }| }^{ 1/2 } }{ { |{ \bm{H} }_{ 0,\bm{\delta} }| }^{ 1/2 } }  \frac { \Gamma (n/2) }{ { {s}_{c} }^{ n/2 } }.
\end{align*}
This is the marginal likelihood \eqref{eq:ML}.

\section{Full conditional posteriors}
Because the prior of ${\delta}_{j}$ is the Bernoulli distribution and the prior of $\omega$ is the beta distribution, it is easy to calculate their full conditional posteriors. 
Here we provide the derivations of the full conditional posteriors of ${\bm{\beta}}_{\bm{\delta}}$ and ${\sigma}^{2}$.

To obtain the full conditional posterior of ${\bm{\beta}}_{\bm{\delta}}$, we first focus on the joint distribution $p(\bm{y}|{\bm{X}}_{\bm{\delta}},{\bm{\beta}}_{\bm{\delta}} ,{ \sigma  }^{ 2 })p({\bm{D}}_{\bm{\beta}}{\bm{\beta}}_{\bm{\delta}} )p({\sigma}^{2})$ given by
\begin{align*}
	&p(\bm{y}|{\bm{X}}_{\bm{\delta}},{\bm{\beta}}_{\bm{\delta}} ,{ \sigma  }^{ 2 })p({\bm{D}}_{\bm{\beta}}{\bm{\beta}}_{\bm{\delta}} )p({\sigma}^{2})\\
	&=\frac { 1 }{ { (2\pi ) }^{ n/2 } {{ \sigma  }^{ 2 }}^{(n+2)/2}}\frac { 1 }{ { (2\pi ) }^{ {p}_{1}/2 }{ |{\sigma}^{2}{ \bm{H} }_{ 0 ,\bm{\delta}}| }^{ 1/2 } } \\
	&\quad\times\exp \left\{ -\frac { 1 }{ 2 } { ({\bm{\beta}}_{\bm{\delta}} -{ \bm{h} }_{ \bm{\delta}  }) }^{ T }{ ({ \sigma  }^{ 2 }{ \bm{H} }_{ \bm{\delta}  }) }^{ -1 }({\bm{\beta}}_{\bm{\delta}} -{ \bm{h} }_{ \bm{\delta}  }) \right\} \exp \left(-\frac { { s }_{ c } }{ { \sigma  }^{ 2 } } \right).
\end{align*}
The terms concerned with ${\bm{\beta}}_{\bm{\delta}}$ can be sorted out as
\begin{equation*}
	\exp \left\{ -\frac { 1 }{ 2 } { ({\bm{\beta}}_{\bm{\delta}} -{ \bm{h} }_{ \bm{\delta}  }) }^{ T }{ ({ \sigma  }^{ 2 }{ \bm{H} }_{ \bm{\delta}  }) }^{ -1 }({\bm{\beta}}_{\bm{\delta}} -{ \bm{h} }_{ \bm{\delta}  }) \right\},
\end{equation*}
which is the kernel of the multivariate normal distribution with mean vector ${ \bm{h} }_{ \bm{\delta}  }$ and variance-covariance matrix ${\sigma}^{2}{ \bm{H} }_{ \bm{\delta}  }$. 
Hence the full conditional posterior of ${\bm{\beta}}_{\bm{\delta}}$ is given by
\begin{equation*}
	{\bm{\beta}}_{\bm{\delta}}|\bm{y},{\sigma}^{2},\bm{\delta}\sim {\rm N}_{p_1+1}({\bm{h}}_{\bm{\delta}},{\sigma}^{2}{\bm{H}}_{\bm{\delta}}).
\end{equation*}

The terms concerned with ${\sigma}^{2}$ can be sorted out as
\begin{equation*}
	{({\sigma}^{2})}^{-n/2-1}\exp \left(-\frac{{s}_{c}}{{\sigma}^{2}} \right),
\end{equation*}
which is the kernel of the inverse Gamma distribution with shape parameter ${n}/{2}$ and scale parameter ${s}_{c}$. 
Hence the full conditional posterior of ${\sigma}^{2}$ is given by
\begin{equation*}
	{\sigma}^{2}|\bm{y},\bm{\delta}\sim {\rm IG} \left( \frac{n}{2},{s}_{c} \right).
\end{equation*}

\bibliographystyle{apalike}
\bibliography{mybibfile}

\end{document}